\documentclass[3p,times,procedia]{elsarticle}
\usepackage{nupha_ecrc}

\volume{00}

\firstpage{1}

\journalname{Nuclear Physics A}

\runauth{D.S.D Albuquerque, for the ALICE Colaboration}

\jid{nupha}

\jnltitlelogo{Nuclear Physics A}

\usepackage{amssymb}

 \usepackage{hyperref}

 \usepackage[figuresright]{rotating}
 \usepackage{xcolor}
 \usepackage{soul}
 \usepackage{lipsum}
 \usepackage[utf8]{inputenc}
 \usepackage{etoolbox}
 \patchcmd{\emailauthor}{(#2)}{}{}{}
 \patchcmd{\urlauthor}{(#2)}{}{}{}

 \usepackage{xspace}
\usepackage{color}
\usepackage{xcolor}
\usepackage{rotating}
\definecolor{light-gray}{gray}{0.8}

\newcommand{\dnchdeta}{\ensuremath{{\rm d}N_{\rm ch}/{\rm d}\eta}\xspace}

\newcommand{\etaless}[1]{\ensuremath{\left|\eta\right| < #1}\xspace}

\newcommand{\dnchdetaless}[1]{\ensuremath{{\rm d}N_{\rm ch}/{\rm d}\eta|_{\etaless{#1}}}\xspace}

\newcommand{\seven}{$\sqrt{s}~=~7$~TeV}

\newcommand{\thirteen}{$\sqrt{s}~=~13$~TeV}

\newcommand{\fivenn}{$\sqrt{s_{\rm{NN}}}~=~5.02$~TeV}
\newcommand{\twosevensixnn}{$\sqrt{s_{\rm{NN}}}~=~2.76$~TeV}
\newcommand{\fivefourfournn}{$\sqrt{s_{\rm{NN}}}~=~5.44$~TeV}

\newcommand{\comment}[1]{}

\begin{document}
 \begin{frontmatter}



	 \dochead{XXVIIth International Conference on Ultrarelativistic Nucleus-Nucleus Collisions\\ (Quark Matter 2018)}

	 \title{Hadronic resonances, strange and multi-strange particle production in Xe-Xe and Pb-Pb collisions with ALICE at the LHC}


	 \author[label1]{D.S.D Albuquerque}
	 \ead{dsilvade@cern.ch}
	 \author{for the ALICE Collaboration}

	 \address{Campinas State University, Brazil}

	 \begin{abstract}
		 We present measurements of hadronic resonance, strange and multi-strange particle production in collisions of Xe-Xe and Pb-Pb at the center-of-mass energies of $\sqrt{s_{NN}}=5.44$ and $5.02$ TeV, respectively, by the ALICE collaboration at the LHC. Particle ratios are presented as a function of multiplicity for $K^{0}_{s}$, $\Lambda$, $\Xi^{-}$, $\bar{\Xi}^{+}$, $\Omega^{-}$, $\bar{\Omega}^{+}$, $\rho(770)^{0}$, $K^{\ast}(892)^{0}$, $\phi(1020)$ and $\Lambda(1520)$. Our results are discussed and compared with predictions of QCD-inspired event generators. Additionally, comparisons with lower energy measurements and smaller systems are also presented.

	 \end{abstract}


 \end{frontmatter}


 \section{Introduction}
 \label{intro}
 The measurement of strange and resonance particle production in relativistic heavy-ion collisions is of great interest to investigate the properties of hadronic matter under extreme conditions. The enhanced production of strange and multi-strange hadrons with respect to non-strange particles was historically considered as one of the signatures of the formation of a partonic phase during the evolution of the system created in such collisions \cite{Rafelski:1982pu}.
Moreover, short-lived hadronic resonances have a lifetime shorter or comparable to the lifetime of the fireball \cite{Markert:2008jc}, and thus can undergo re-scattering and regeneration processes \cite{Bleicher:2002dm,Bleicher:2003ij}, causing their yield to deviate from expected thermal model predictions \cite{Wheaton:2004qb,Andronic:2008gu,Petran:2013dva,Stachel:2013zma}. Therefore, the study of resonances with different lifetimes is interesting to characterize the late stages of the fireball evolution.
The ALICE detector is especially suited for the study of the Quark-Gluon Plasma (QGP) via relativistic heavy-ion collisions thanks to its exceptional particle identification capabilities. Recently, a systematic study was performed with a wide range of strange particles and hadronic resonances being measured in different colliding systems and at various energies, including the new Xe-Xe data sample collected in late 2017.
The full description of the ALICE detector can be found in \cite{Aamodt:2008zz,Abelev:2014ffa} and for further information regarding the techniques used for the measurements reported here see \cite{Abelev:2013xaa,ABELEV:2013zaa,Abelev:2014uua,ALICE:2018ewo,Acharya:2018qnp}.

 \section{Strangeness Enhancement}

 The enhancement of strange particles in heavy-ion collisions was historically quantified by calculating the strange particle yield per participant pair, as estimated with the Glauber model \cite{Abelev:2013qoq},  and comparing the result to the values obtained in proton-proton (pp) collisions at the same center-of-mass energy \cite{ABELEV:2013zaa}. In recent results, we see that this might not be the best approach, since the production rate of strange particles in pp collisions also increases with the event multiplicity \cite{ALICE:2017jyt}.

 \begin{figure}[h!]
 	\centering
	\includegraphics[width=0.55\textwidth]{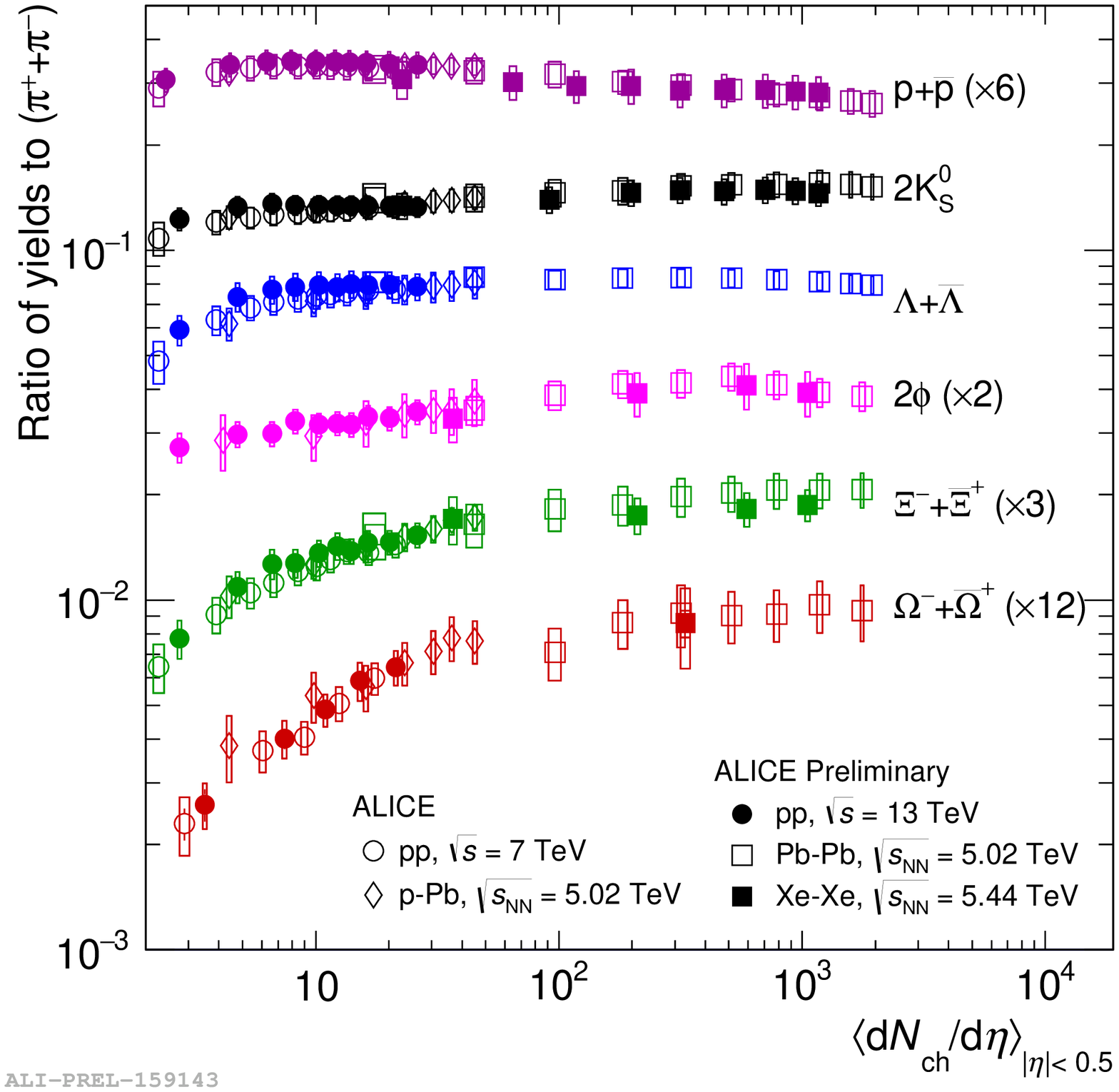}
	\caption{Ratio to pion of integrated yields for $p$, $K_{s}^{0}$, $\Lambda$, $\phi$, $\Xi$ and $\Omega$. The evolution with multiplicity at mid-rapidity, \dnchdetaless{0.5}, is reported for several systems and energies, including pp at \seven\ \cite{ALICE:2017jyt}, p-Pb at \fivenn\ \cite{Adam:2015vsf}, and also the ALICE preliminary results for pp at \thirteen, Xe-Xe at \fivefourfournn\ and Pb-Pb at \fivenn\ are included for comparison. Error bars show the statistical uncertainty, whereas the empty boxes show the total systematic uncertainty.}
	\label{fig:StrangenessSummaryPlot}
 \end{figure}
 
 A better way to quantify this enhancement would be to normalize the total yield of strange particles to the yield of pions. To compare the relative increase of strange particles across different colliding systems, the yield ratios are plotted as a function of charged particle multiplicity. In figure \ref{fig:StrangenessSummaryPlot}, the evolution of $p$, $K_{s}^{0}$, $\Lambda$, $\phi$, $\Xi$ and $\Omega$ yield ratios with respect to pions is reported as a function of multiplicity for all the available colliding systems.

 We observe a smooth evolution with multiplicity across all systems, from low multiplicity pp to central AA collisions. The enhancement is observed to be more pronounced for particles with a larger strangeness content, with the $\phi$ exhibiting a behavior that is intermediate between $K^{0}_{s}$ and $\Xi$. A saturation value seems to be reached in AA collisions for $\langle \dnchdeta \rangle_{|\eta|<0.5} \gtrsim100$, in close agreement with predictions from statistical hadronization models (SHMs) \cite{Stachel:2013zma}. Ultimately, the new measurements in Xe-Xe collisions further corroborate the observation that, for a given particle species and multiplicity, particle ratios are independent of energy and system.

 \section{Resonance yields and properties of the hadronic phase}

 The measured hadronic resonance yields may be influenced by several factors: initial yield at chemical freeze-out, resonance lifetime, scattering cross-section of resonance decay daughters and lifetime of the hadronic phase of the system evolution. Therefore, a comparison of measured resonance yield to the production rate of its stable counterpart can provide information about the late stage of the system evolution. Hadronic resonances reported here are reconstructed via the detection of their strong-decay daughters. The lifetimes of the measured resonances vary significantly, as indicated in table \ref{tab:ResonanceTable}.

 \begin{table}[h!]
	 \centering
	 \begin{tabular}{ccccc}
	 						& $\rho(770)^0$ & $K^{\ast}(892)^{0}$ & $\Lambda(1520)$ & $\phi(1020)$ \\ \hline
\textbf{Lifetime ($\mathrm{fm}/c$) }& 1.3           & 4.2                 & 12.6            & 46           \\
\textbf{Decay mode}                 & $\pi^{+}\pi^{-}$      & $K^{+}\pi^{-}$              & $pK^{-}$            & $K^{+}K^{-}$         \\
\textbf{B.R. ($\%$)}                & $~100$        & $~100$              & $22.5$          & $48.9$      
     \end{tabular}
	 \caption{Lifetime values, reconstructed decay mode and branching ratio for $\rho(770)^0$, $K^{\ast}(892)^{0}$, $\Lambda(1520)$ and $\phi(1020)$.}
     \label{tab:ResonanceTable}
 \end{table}

\begin{figure}[h!]
 	\centering
		\includegraphics[width=0.75\textwidth]{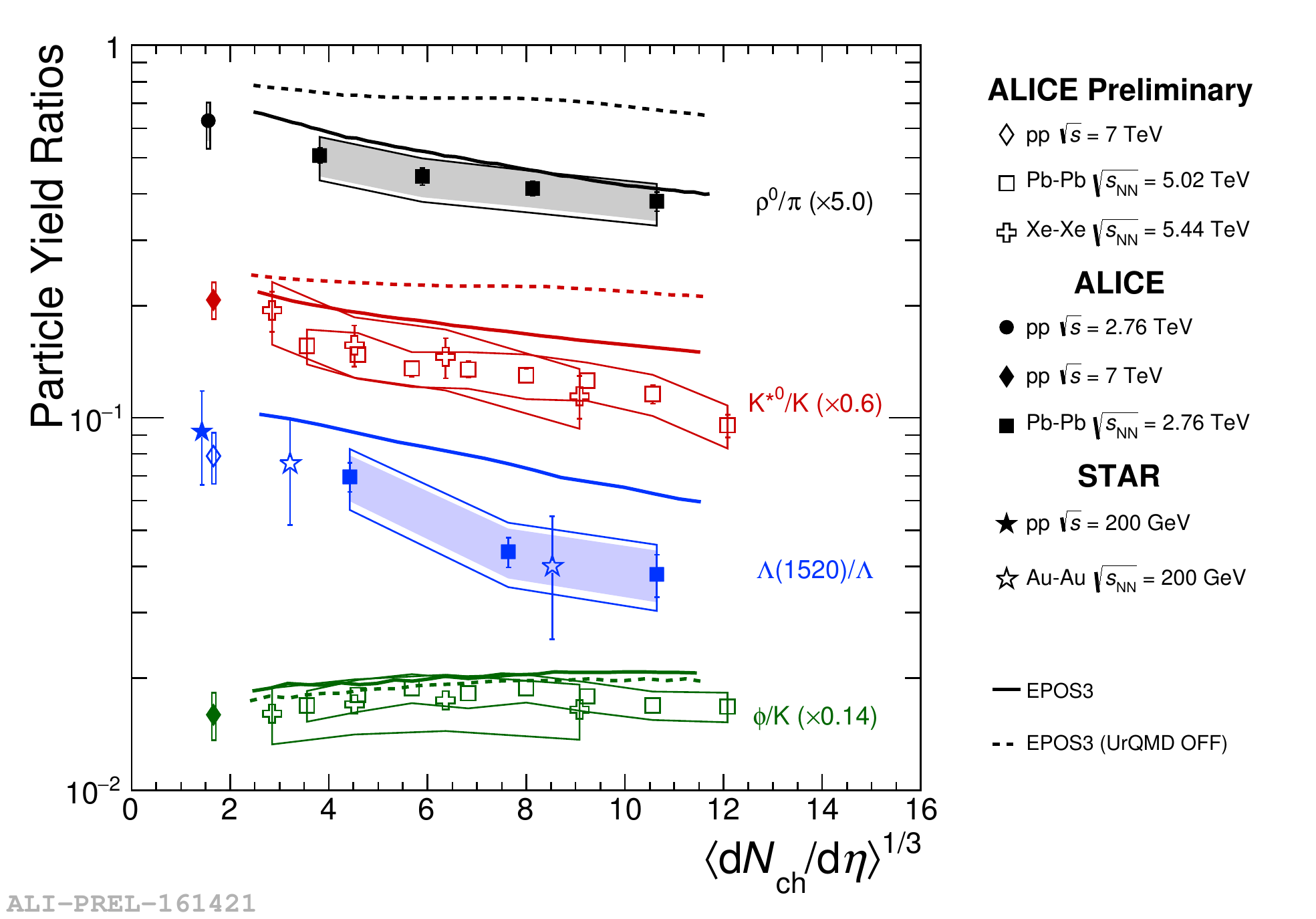}
	\caption{Ratio of resonance and stable counterpart yields is plotted as a function of multiplicity for the hadronic resonances, $\rho^{0}/\pi$ \cite{Acharya:2018qnp}, $K^{\ast0}/K$, $\Lambda(1520)/\Lambda$ \cite{ALICE:2018ewo} and $\phi/K$ from INEL$>0$ pp, Pb-Pb and Xe-Xe collisions. The error bars show the statistical uncertainty, while the empty and dark-shaded boxes show the total systematic uncertainty and the uncorrelated contribution across multiplicity bins, respectively. A comparison to model is also reported by showing the ratios predicted by the EPOS3 model for Pb-Pb at \twosevensixnn, where the late hadronic cascading is simulated with UrQMD. STAR data are also shown for the $\Lambda(1520)/\Lambda$ ratio \cite{Adams:2006yu}.}
	\label{fig:ResonanceSummaryPlot}
 \end{figure}

 The ratio of integrated yields is shown in figure \ref{fig:ResonanceSummaryPlot} for $2\rho(770)^{0}/(\pi^{+}+\pi^{-})$, $(K^{\ast}(893)^{0}+\bar{K}^{\ast}(893)^{0})/(K^{+}+K^{-})$, $(\Lambda(1520)+\bar{\Lambda}(1520))/(\Lambda+\bar{\Lambda})$ and $2\phi(1020)/(K^{+}+K^{-})$, henceforth denoted by the shorthand notation $\rho^{0}/\pi$, $K^{\ast0}/K$, $\Lambda(1520)/\Lambda$ and $\phi/K$, for pp, Pb-Pb and Xe-Xe collisions.
Short-lived resonances show a sizable dependence on multiplicity. A clear suppression is observed for $\rho^{0}$ going from pp to Pb-Pb collisions at \twosevensixnn\ \cite{Acharya:2018qnp}. This evolution with multiplicity is well described by the EPOS3 event generator \cite{Drescher:2000ha,Werner:2010aa,Werner:2013tya}, which includes UrQMD \cite{Bass:1998ca,Bleicher:1999xi} for the late stage hadronic cascading. The attempt to describe the $\rho^{0}/\pi$ ratio without the UrQMD component of the evolution fails to reproduce the data, with an overestimated and multiplicity-independent $\rho^{0}/\pi$ ratio \cite{Knospe:2015nva}. 

 The same behavior is observed in the $K^{\ast0}/K$ ratio, where the suppression can be observed in a wider multiplicity range with data from pp at \seven\ \cite{Abelev:2012hy}, Pb-Pb at \fivenn\ and Xe-Xe at \fivefourfournn. Predictions from EPOS3 describe the evolution of particle ratios with multiplicity.
 
 Despite having a lifetime almost ten times longer than the $\rho^{0}$, new measurements of the $\Lambda(1520)$ production in Pb-Pb reveal that it is suppressed with respect to the $\Lambda$ \cite{ALICE:2018ewo}, similarly to what is observed for other short-lived resonances. However, at present, EPOS predictions can only qualitatively describe the data, showing a similar decrease with multiplicity but with a higher $\Lambda(1520)$ yield. It is also worth noting that the results are in agreement with previous measurements by STAR \cite{Adams:2006yu}.

 Finally, there is small dependence of the $\phi/\pi$ ratio with multiplicity for all measured systems. This is expected in the context of re-scattering, considering that the $\phi$-meson lives longer than the expected fireball lifetime and therefore its decay daughters will not undergo re-scattering.

 \section{Conclusions}

 Latest results on strange, multi-strange and resonances in heavy-ion collisions are reported here, including new results from the Xe-Xe collisions at \fivefourfournn. Comparing the yield of strange particles to the yield of pions, an enhancement of strange particle production is observed as the event multiplicity increases. The enhancement is more pronounced for particles with higher strangeness content and, for a given multiplicity, there is no dependence with both the colliding system and energy. 
 
 The measurement of hadronic resonances is also performed for various systems (from pp to Pb-Pb, including Xe-Xe) and one can observe that, similarly to the ground state, particle ratios seem to depend solely on the final charged-particle density for all systems. A comparison of the resonance suppression with EPOS3 prediction indicates that re-scattering and regeneration processes are required to describe the measurements.

 \section{Acknowledgement}

 This work was supported by CNPq (Conselho Nacional de Desenvolvimento Científico e Tecnológico) grant number 141186/2015-1.




\bibliographystyle{utphys}
\bibliography{My_Collection.bib}







\end{document}